\documentclass[twocolumn,fleqn,usenatbib,12pt,tighten]{mnras}

\usepackage{newtxtext,newtxmath}

\usepackage[T1]{fontenc}

\DeclareRobustCommand{\VAN}[3]{#2}
\let\VANthebibliography\thebibliography
\def\thebibliography{\DeclareRobustCommand{\VAN}[3]{##3}\VANthebibliography}

\usepackage{graphicx}	
\usepackage{amsmath}	

\usepackage{times}
\usepackage{rotating,lscape}
\usepackage{epsfig}
\usepackage{graphicx}
\usepackage{epstopdf}

\usepackage{xcolor}
\usepackage{booktabs}
\usepackage{amsmath}
\usepackage{color,soul}
\usepackage[flushleft]{threeparttable}

\def\chandra{{\it Chandra}~}

\def\lya{\ifmmode {\rm Ly}\alpha~ \else Ly$\alpha$~\fi}
\def\lyan{\ifmmode {\rm Ly}\alpha \else Ly$\alpha$\fi}
\def\lyb{\ifmmode {\rm Ly}\beta~ \else Ly$\beta$~\fi}
\def\lyg{\ifmmode {\rm Ly}\gamma~ \else Ly$\gamma$~\fi}
\def\civ{\ifmmode {\rm C}\,{\sc iv}~ \else C\,{\sc iv}~\fi}
\def\civn{\ifmmode {\rm C}\,{\sc iv}~ \else C\,{\sc iv}\fi}
\def\cvi{\ifmmode {\rm C}\,{\sc vi}~ \else C\,{\sc vi}~\fi}
\def\cvin{\ifmmode {\rm C}\,{\sc vi} \else C\,{\sc vi}\fi}

\def\nv{N\,{\sc v}~}

\def\siii{Si\,{\sc II}~}

\def\siiiin{Si\,{\sc III}}

\def\sixiin{Si\,{\sc XII}}

\def\arxiiin{Ar\,{\sc XIII}}

\def\arxivn{Ar\,{\sc XIV}}
\def\caxi{Ca\,{\sc XI}~}

\def\oii{{{\rm O}\,{\sc ii}~}}

\def\ovi{{{\rm O}\,{\sc vi}~}}

\def\ovii{{{\rm O}\,{\sc vii}~}}
\def\oviii{{{\rm O}\,{\sc viii}~}}

\def\oviin{{{\rm O}\,{\sc vii}}}

\def\nex{{{\rm Ne}\,{\sc x}~}}

\def\msun{{M$_{\odot}$}}
\def\kms{{km s$^{-1}$}}

\def\H18{{\it H1821+643}~}

\title[Hot CGM in X-ray absorption]{Probing the hot circumgalactic medium of external galaxies in X-ray absorption II: a luminous spiral galaxy at $z\approx 0.225$
}

\author[Smita Mathur et al.]{
Smita Mathur,$^{1,2}$\thanks{E-mail: mathur.17@osu.edu}
Sanskriti Das,$^{3}$
Anjali Gupta$^{4}$
and Yair Krongold$^{5}$
\\
$^{1}$Astronomy Department, The Ohio State University, Columbus,
  OH 43210, USA\\
$^{2}$Center for Cosmology and
  Astro-Particle Physics, The Ohio State University, Columbus, OH 43210\\
$^{3}$ Kavli Institute for Particle Astrophysics and Cosmology, Stanford University, 452 Lomita Mall, Stanford, CA\,94305, USA\\
$^{4}$Columbus State Community College, Columbus, OH, USA \\
$^{5}$ Instituto de Astronomia, Universidad Nacional Autonoma de Mexico, 04510 Mexico City, Mexico \\
}
\date{Accepted XXX. Received YYY; in original form ZZZ}

\pubyear{2023}

\begin{document}
\label{firstpage}
\pagerange{\pageref{firstpage}--\pageref{lastpage}}
\maketitle

\begin{abstract}
The circumgalactic medium (CGM) is the most massive baryonic component of a spiral galaxy, shock heated to about $10^6$K for an $\rm L^{\star}$ galaxy. The CGM of the Milky Way has been well-characterized through X-ray absorption line spectroscopy. However, the paucity of bright background sources makes it challenging to probe the CGM of external galaxies. Previously, using broad \ovi absorption as a signpost, we successfully detected the CGM of one galaxy  in X-rays. Here we report on the detection of the \ovii $K\alpha$ absorption line at the redshift of a spiral galaxy at $z\approx0.225$ using 1.2\,Ms of \chandra  observations. This is a robust detection, clearly showing the presence of the hot gas. The mass in the hot phase is at least an order of magnitude larger than that in the cooler phases detected in the UV. The presence of hot gas $116h^{-1}$kpc from the center of this galaxy provides credence to the existence of the  extended CGM of the Milky Way. There has been a report of the detection of \ovii absorption from the warm-hot intergalactic medium in this sightline using stacking analysis on older dataset. We argue that the absorption line is from the CGM of the $z\approx0.225$ galaxy instead.
\end{abstract}




\section{Introduction}

The circumgalactic medium (CGM) is the gaseous medium filling the halo of a spiral galaxy. It is the most volume filling and the most massive baryonic component of a galaxy, shock heated to about $10^6$K for an $\rm L^{\star}$ galaxy, in the process of galaxy formation. Therefore, it is critical to understand the CGM to understand galaxy formation and evolution (see the review by \citet{Tumlinson2017,Faucher-Giguere2023}). Given the high temperature, the CGM can be observed effectively in X-rays (see the review by \citet{Mathur2022}).

At high temperatures ($\approx
10^6$--$10^7$K) metals become highly ionized, dominated by He-like and H-like charge states,
observable only through X-ray absorption and emission lines; we will refer to this temperature range as ``warm-hot'' or ``hot" in this paper. 
The Milky Way CGM has been studied in both X-ray emission and absorption, providing complementary information. The emission observations \citep{Snowden2000,Smith2007,Galeazzi2007,Henley2007,Henley2008,Gupta2009,Das2019c,Gupta2023,Bhatt2023} are biased to detecting denser gas at a narrow temperature range where the  emissivity of the ion (say, \oviin) peaks. Absorption studies, on the other hand, are necessary to detect low-density diffuse gas permeating the halo, over the entire temperature range over which the ionization fraction of an ion (say, \oviin) is high. Absorption line spectroscopy has been used effectively to characterize the Milky Way CGM \citep{Nicastro2002,Williams2005,Fang2006,Gupta2012,Gupta2014,Miller2015,Das2019a,Das2021b,Lara2023}. 

Detecting the hot CGM of individual external galaxies, however, is difficult. For emission studies, the CGM flux drops with distance from us, making it hard to isolate the CGM signal from the background. It is no surprise, therefore, that the CGM of a few ($\sim 8$) super-$L^{\star}$ galaxies and only one $\approx \rm L^{\star}$ galaxy, NGC\,3221, have been detected in emission \citep[and references therein]{Das2019b,Das2020a}.  Absorption line spectroscopy poses different challenges. Given the limitations of current technology, the background sources need to be very bright to obtain a good enough spectrum for detecting weak intervening absorption lines. Such X-ray bright quasars are few, and not necessarily well-aligned with foreground galaxies, making blind searches of intervening absorbers unfeasible. 

In order to successfully detect $z>0$ X-ray absorption lines, we can use "signposts" of absorption systems at other wavelengths.  Typically, \ovi absorption lines are produced in gas at $T\approx 10^{5.5}$K. However, if they are produced in the hot gas, the line would be thermally broadened. This makes broad \ovi absorption lines a good proxy for the hot gas. \citet{Mathur2021} observed the sightline to PKS$0405-123$ passing through the CGM of a galaxy at $z\approx 0.167$, with known broad \ovi absorption. They detected the \ovii $K\alpha$ absorption line at the galaxy redshift and their hot phase model was consistent with that predicted from UV data by \citet{Savage2010}. Alternatively, one can use Lyman limit systems (LLSs) as signposts of intervening galaxies. \citet{Nicastro2023} stacked X-ray spectra of three background quasars with known LLSs and successfully detected \ovii absorption associated with the CGM of the galaxies. 
Here we report on the detection of the \ovii absorption line at the redshift of an intervening \ovi absorption system in the sightline to an X-ray bright blazar. 

\subsection{Sightline to \H18}

The sightline to the UV- and X-ray-bright blazar \H18 has been studied extensively. There are six intervening \ovi absorption systems in this sightline, observed with the Hubble Space Telescope (HST; \citet{Tripp2000}) and the Far Ultraviolet Space Explorer (FUSE; \citet{Oegerle2000}). We observed \H18 with \chandra Low Energy Transmission Grating (LETG) and  Advanced CCD Imaging Spectrometer (ACIS) for 500ks to search for the elusive warm-hot intergalactic medium (\citet{Mathur2003} at the redshifts of the \ovi systems ($z1=0.26659, z2=0.24531, z3=0.22637, z4=0.22497, z5=0.21326,
\rm{and}~ z6=0.12137$). The high redshift intergalactic medium (IGM) is traced by the \lya forest, and most of the baryonic mass in the high redshift Universe resides in the IGM. At lower redshifts ($z\leq 2$) the IGM is shock-heated in the process of large-scale structure formation, but most of the low redshift baryons are believed to remain in the warm-hot IGM (WHIM; \citet{Dave2010}). 
\citet{Mathur2003} reported marginal ($2\sigma$) detections of \ovii at $z2, z4$ and $z6$; 
 of \oviii at $z6$, and of  \nex at $z4$. \citet{Kovacs2019} used the same \chandra LETG-ACIS-S data and performed stacking analysis to look for \ovii absorption at the redshifts of 17 \lya absorption systems. They reported significant ($3.3\sigma$) detection of the \ovii line and argued that the line originates from the WHIM. One of the 17 systems in their sample, at $z\approx 0.225$, has the smallest impact parameter of 112.5kpc to an intervening galaxy; they do not discuss whether the \ovii absorption could be from the CGM of this galaxy. Previously, the same absorption system was studied extensively by \citet{Narayanan2010}. They found that the two \ovi absorption systems, at $z=0.22496$ and $z=0.22638$, are associated with the extended halo of the same spiral galaxy situated at a projected distance of $116h^{-1}_{71}$kpc (see also \citet{Savage1998}).  In this paper, we present new 1.2\,Ms \chandra observations of the \H18 sightline and show that the \ovii absorption arises in the hot CGM of the $z\approx 0.225$ galaxy. 

\section{Observations and analysis}

New \chandra observations of \H18 were made from March, 2020 through August, 2021. The source was observed with LETG/HRC-S (High Resolution Camera for Spectroscopy) for 1.2Ms (sequence numbers 601449 and 601495). We analyzed these data using CIAO (Chandra Interactive Analysis of Observations), following standard data reduction tools and threads\footnote{\url{https://asc.harvard.edu/ciao/}}. Since HRC-S does not have clear order sorting, we included grating orders 1--8 in our analysis and combined the plus and minus orders. The spectrum showed practically no data beyond 3 keV, so we analyzed the spectrum in the 0.3--3 keV range. The observed net count rate was  $0.2485\pm0.0005$ counts/s yielding a total of $2.98\times 10^5$ background-subtracted counts. 
 We binned the spectrum by four, to the LETG/HRC-S resolution of  0.05\AA; the absorption lines are not resolved with \chandra LETG. We also note that the wavelength calibration uncertainty of LETG/HRC-S has the mean value of about 0.01 \AA\ (120 \kms at 25\AA) with a large scatter around the mean\footnote{\url{ https://cxc.harvard.edu/cal/Letg/Hrc\_disp/}}. All the spectral analysis was performed with XSPEC v12.12.1\footnote{\url{https://heasarc.gsfc.nasa.gov/xanadu/xspec/}}. The errors are $1\sigma$, unless noted otherwise. 

As noted above, there are six \ovi absorption line systems in the sightline to \H18. The \ovii K$\alpha$ line at the rest wavelength of 21.602\AA\ is expected to be the strongest line in hot plasma (see \citet{Mathur2003,Mathur2022}), which is the focus of WHIM or CGM studies. The \ovii K$\alpha$ line is redshifted to 27.3608\AA\ for z1, 26.9012 \AA\ for z2, 26.4920 for z3, 26.4618 for z4, 26.2088 for z5, and 24.2238 for z6. The focus of this paper is the CGM of the galaxy associated with the z4 system at $z=0.22497$ and the z3 system at $z=0.22637$. The z4 \ovi absorber is significantly stronger (N(\ovi)$=19.9\pm 1.2 \times 10^{13}$ cm$^{-2}$) than z3 (N(\ovi)$=2.4\pm 0.5 \times 10^{13}$ cm$^{-2}$); we will refer to this absorption system as the ``z4 system'' and the associated galaxy as the ``z4 galaxy''  for the purpose of this paper. The observed wavelength difference of 0.0302 \AA\ corrsponding to the z3 and z4 systems is less than the LETG/HRC-S resolution, thus the two \ovii lines would be blended in the \chandra spectrum. 

The 0.3--3.0 keV spectrum of \H18 cannot be simply fit with an absorbed power-law \citep{Mathur2003}. We note that the observed \ovii wavelengths of all the six absorption systems are within the 24--28 \AA\ range, therefore we restricted our analysis to this range. This enables us to fit the local continuum with a simple power law, as also done in \citet{Lara2023}. We added Gaussian lines at the locations of the wavelengths of the expected \ovii lines, which were held fixed. The lines are not resolved at the LETG resolution, therefore the line width was held fixed at a very small value of $10^{-5}$; the observed line width is thus the instrumental line width. The free parameters of the initial spectral fit were Galactic column density N(H), power-law slope and normalization, and the normalization of each Gaussian line. The result of this fit is shown in Figure 1. The only features clearly seen in this spectral range are the line corresponding to the "z4 system", and a marginal line corresponding to the z2 system.  

In Figure 2, we show the same fit, but the ``unfolded" spectrum is plotted instead, where the data are deconvolved  by the effective area of the instrument. This is akin to a flat-fielded spectrum in optical/UV. Here the z4 and z2 systems are clearly seen. In Figure 3 we show the normalized spectrum from 26\AA\ to 27\AA, around the z4 line. The flux at the z4 line center is $2.78\pm 0.33 \times 10^{-4}$ photons cm$^{-2}$ s$^{-1}$ \AA$^{-1}$ and the continuum flux at the same wavelength is $4.30 \pm 0.97 \times 10^{-4}$ photons cm$^{-2}$ s$^{-1}$ \AA$^{-1}$. Thus the flux at the line trough is $4.48\sigma$ below the continuum.  

A conventional way to assess the presence of an additional spectral feature is done through the F-test.  We obtain the best-fit $\chi^2=45.01$ for 71 degrees of freedom including the z4 line. Without the line, we had $\chi^2=53.89$ for 72 degrees of freedom. Thus the F-test probability for finding the line feature by chance is $0.05\%$. This shows that the line is present at 99.95\% confidence. The line equivalent width, however, is measured with a lower significance of $3\sigma$: $\rm EW=0.38\pm 0.13$eV ($21 \pm 7$m\AA). However, it is very important to note that all these measures of significance or confidence (derived by any of these three methods) are not for blind searches, but for targeted searches with the \ovi prior. Thus even a $3\sigma$ result is a robust result.


Next, we allowed the line wavelength to be free within 0.1\AA\ (two resolution elements) around the expected z4 line center (26.4618 \AA) and refitted the spectrum. The best-fit line centroid is $26.4385^{+0.009}_{-0.012}$\AA. The best-fit wavelength is thus within half the resolution element of the expected wavelength. 

While the focus of this paper is the z4 system, we note the absorption line at $26.9012$ \AA\ from the z2 system. The line is marginally detected but is required by the fit. If we start with the line normalization of zero and re-fit the spectrum, the best-fit normalization is non-zero. The best-fit $EW=10.7+^{6.3}_{7.3}$m\AA. The same line is also reported by \citet{Mathur2003} with $EW=13.9\pm6.2$m\AA\ consistent with what we find here. \citet{Kovacs2019} note a $z=0.24568$ galaxy at an impact parameter of $669.6$kpc. Given the large impact parameter, it is unlikely that the observed absorption is from the CGM of that galaxy. There might be a closer but fainter galaxy or the absorption feature may not be real, or it may actually arise from the WHIM (see \citet{Mathur2003} for details); we have no further information to choose among these options. We do not discuss this z2 system further. 
 
\subsection{Combined analysis of LETG- HRC-S and -ACIS data}

To obtain better constraints on the \ovii absorption line at z4, we reanalyzed the LETG-ACIS-S data \citep{Mathur2003} and fitted the 26--27 \AA\ spectrum simultaneously with the LETG-HRC-S spectrum. The measured EW of the \ovii line is $EW=0.3\pm0.1$eV ($16.95\pm 5.65$ m\AA), consistent with the measurement using the LETG-HRC-S spectrum only.  This is the observed EW at z4; the rest-frame EW at the z4 galaxy is, therefore, $EW=13.8\pm 4.6$m\AA. Following equation 2 in \citet{Mathur2003}, we calculate the \ovii column density to be $\rm N(O VII)=4.9\pm 1.6 \times 10^{15} cm^{-2}$. 

\section{Discussion}

\subsection{A z=0 line?}

In the above section, we showed that a line is clearly present at 26.4618\AA. Assuming this is to be \ovii K$\alpha$, we argue that it arises in the CGM of the ``z4 galaxy''.  However, could it be another ionic transition at another redshift? In particular, we need to check whether the observed line feature could be a z=0 line from the interstellar medium (ISM) or the CGM of the Milky Way. Indeed, \citet{Mathur2017} showed that the putative $z=0$ \ovi line at $\lambda=22.03$\AA\ from the Milky Way CGM is in fact an \oii line at $\lambda=22.04$ \AA\ from the ISM, providing a solution to $\approx 15$ year mystery of the mismatch between the X-ray and UV measured \ovi column densities. Recently, \citet{Gatuzz2023} argued that the putative intervening \ovii absorption lines from the WHIM reported in literature along five different sightlines are in fact lower-ionization lines from the interstellar medium of our Galaxy. Therefore, we searched for the line transitions of C, N, O, Ne, Mg, and Si with rest-wavelengths in the $26-27$\AA\ range following \citet{Gatuzz2023} and references therein. \citet{Gatuzz2021} report a \nv K$\gamma$ line around $26$\AA. However, a stronger \nv $K\alpha$ line at $29.4135$\AA\ is not detected in our spectrum. Therefore we rule out that the observed line is a $z=0$ \nv line. No line from any other element is present in the wavelength range of interest. 

 We next searched NIST\footnote{\url{https://www.nist.gov/pml/atomic-spectra-database}} for lines from any other element in the 26--27\AA\ range. There are possible weak lines of \sixiin, \arxiiin, \arxivn, and \caxi in this range, but the stronger lines of these ions are not detected in our spectrum. We thus rule out that our observed line at $\lambda=26.4618$\AA\ is any z=0 line. 

\subsection{The CGM of the $z\approx 0.225$ galaxy}

 \citet{Narayanan2010} studied the CGM of the z3/z4 galaxy extensively in the UV. They found that the \ovi absorption line in the z4 system is broader than the photoionized lower ions such as \siii and \siiiin. They concluded that \ovi likely originates in collisionally ionized plasma under non-equilibrium conditions. 
They proposed that the observed \ovi absorber is analogous to Galactic \ovi high-velocity clouds, produced in the transition region between the warm phase and the hot ($\rm T >10^6$K) corona of the galaxy. The \ovi absorber at z3 has an associated broad \lya absorber.  \citet{Narayanan2010} conclude that the z3 system traces a cooling condensing fragment in the galaxy’s hot gaseous halo. Thus, both the z3 and z4 \ovi systems predict the presence of the hot CGM of the host galaxy at $z=0.2256$. 

In this paper, we show that the z4 galaxy indeed harbors hot CGM as traced by the \ovii absorption line and the \ovii column density is $\rm N(O VII)=4.9\pm 1.6 \times 10^{15} cm^{-2}$. Assuming oxygen abundance to be $\rm A_O/A_H=8.51\times 10^{-4}$ \citep{Asplund2009},  metallicity of $Z/Z_{\odot}=0.1$ \citep[for consistency with][]{Narayanan2010}, and ionization fraction of \ovii in the hot CGM to be $f_{OVII}=0.5$ \citep[following][]{Gupta2012}, we calculate the total column density.
\begin{equation}
    \rm N_H=1.1\times 10^{20}\;cm^{-2} \Big(\frac{A_O/A_H}{8.51\times 10^{-4}}\Big)^{-1} \Big(\frac{Z/Z_{\odot}}{0.1}\Big)^{-1} \Big(\frac{f_{OVII}}{0.5}\Big)^{-1}  
\end{equation}

The \H18 sightline passes at $d=116$ kpc from the galaxy.  \citet{Narayanan2010} have assumed the radius of the halo of the galaxy $R=200$kpc, and we use the same for the sake of consistency. Thus the pathlength traced by the \ovii absorber through the CGM is $\rm L=2 \times \sqrt{(R^2 - d^2)} = 325.8$ kpc. Given $\rm N_H= \mu n_e L$, where $\mu=0.8$ is the mean molecular weight, we calculate the density $n_e=1.4 \times 10^{-4} cm^{-3}$.   The density we measure is the average density along the pathlength, between the radii $d$ and $R$ of the galaxy. For any density profile monotonically decreasing with radius, the density would be higher at radii $r<d$ and lower for radii $r>d$.  Thus the measured density is a conservative estimate of the average density of the entire halo. Using this density, a (conservative) constant density profile, and the spherical distribution of the hot gas filling the halo out to the radius R, we calculate the mass of the hot gas in the CGM. 
\begin{equation}
\rm M_{CGM}= 9.1\times 10^{10}\;M_\odot \Big(\frac{R_{200}}{200kpc}\Big)^2 \Big(\frac{N_H}{1.1\times 10^{20} cm^{-2}}\Big)
\end{equation}

Thus the mass in the hot CGM is over an order of magnitude larger than the mass probed the UV systems ($6\times 10^9$\msun; \citet{Narayanan2010}). 


The stellar mass of the Milky Way is $4-6 \times 10^{10}$\msun\ and the total mass is $1-2 \times 10^{12}$\msun. We do not know the mass of the $z\approx 0.225$ galaxy, but it is a Milky Way-like galaxy (given $R=200$kpc noted above); therefore we will assume minimum $M_\star=4\times 10^{10}$\msun\ and total mass $M_{total}= 10^{12}$\msun. The total baryonic mass of the galaxy is $M_b=M_{\star}+M_{CGM}=13\times 10^{10}$\msun, ignoring contributions from the ISM and the cooler phases of the CGM. Thus the baryon fraction we measure is $f_b=0.13$, compared to the cosmological $f_b=0.157$ \citep{Planck2016}. 
Note that our measurement has several uncertainties: (1) the measured density is the average density beyond 116 kpc, with the true average being larger (depending on the density profile);  (2) we have not included saturation effects \citep{Williams2005,Gupta2012}; and (3) metallicity is assumed to be $0.1 Z_{\odot}$. The first two of these have likely lowered the measured baryonic mass, while the third has increased it. 
Without knowing the density profile and the metallicity we cannot calculate the exact total baryonic mass of the halo and without knowing the stellar mass and halo mass we cannot determine the actual baryon fraction of the galaxy, but the calculations presented here show that the numbers are reasonable.

Alternatively, we estimate the stellar mass of the galaxy as follows.  \citet{Narayanan2010} have published the B magnitude of the galaxy, but in order to determine the stellar mass we need infrared magnitudes. 
We obtained the WISE W1 magnitude at $3.4\mu m$  from the NASA/IPAC Science Archive (IRSA), from the AllWISE Source Catalog\footnote{
https://irsa.ipac.caltech.edu}. From W1=14.306, we calculated the galaxy luminosity to be 
$\rm \nu L_{\nu} = 8 \times 10^{43} erg~ s^{-1}$ (assuming a $\Lambda$CDM cosmology with $\rm H_0=69.6, \Omega_m = 0.286 ~and~ \Omega_{Lambda}=.714$). We used the stellar mass--$3.4\mu m$ luminosity correlation of \citet{Wen2013} to calculate the stellar mass of the galaxy $\rm M_{\star}=2.2\times 10^{11}$\msun. \citet{Coupon2015} have determined the stellar mass--halo mass correlation using joint analysis with clustering, lensing, and abundance matching. Accordingly, we determine the halo mass of our target galaxy to be $\rm M_h=5\times 10^{13}$\msun. From this halo mass, we calculated the virial radius of the galaxy $R_{200}=783$kpc. Thus we see that both the mass and the radius is huge, almost in the range of groups of galaxies. In the density calculation above, we had assumed the galaxy radius to be 200 kpc. If instead it is 783 kpc, the density would be $\rm \log n_H (cm^{-3})\approx -4.33$. The mass in the hot CGM would then be larger by a factor of $(\frac{R_{200}}{200kpc})^2= 15.3$, making $\rm M_{CGM}= 1.4\times 10^{12}$\msun. Once again, the mass in the hot CGM is significantly higher than the stellar mass of the galaxy. The baryon fraction, however, would be significantly smaller $f_b=0.03$.

Hydrodynamic simulations of \citet{Oppenheimer2016} predict temperature--density phase space for a similar halo, with $M_h=1.6\times10^{13}$\msun\ and $R_{200}=486$ ~kpc (their fig. 3). While there is a small distribution around the mean, the temperature peaks at $\rm \log T (K)=6.4-6.5$, and density at $\rm \log n_H (cm^{-3})\approx -4.25$. The temperature is in the range where \ovii ionization fraction peaks and the density we calculate is within their predicted range. This shows that our estimates are consistent with those from theoretical simulations.

The above sets of calculations show the huge uncertainty in the derived parameters. We also note that there is an order of magnitude scatter in the stellar mass--$3.4\mu m$ luminosity correlation of \citet{Wen2013}, thus the stellar mass of our target galaxy could be significantly smaller. The halo mass is also a strong function of stellar mass \citep{Coupon2015}, therefore the baryon fraction would also likely depend on the stellar mass, which itself is highly uncertain. Once again, the calculations presented here show how the derived parameters depend on assumptions and how uncertain they are.


\subsection{Relevance for the CGM studies of the Milky Way}

Because of the special vantage point, the CGM of the Milky Way has been studied in detail with X-ray absorption lines (see the review by \citet{Mathur2022}). However, because of our location in the disk of the Galaxy, we cannot directly measure the location of the absorbing gas, or its pathlength. Are the observed $z=0$ absorption lines truly from the CGM or could they arise in the interstellar medium or the extra-planar region of the Galaxy? By combining absorption and emission observations we showed that the MW CGM is diffuse, extended to over 100 kpc, and massive \citep{Gupta2012}. There is no anti-correlation of the absorption column density with Galactic latitude $sin(b)$ of the sightline \citep{Gupta2014,Fang2015}, which would have been present if the absorption was dominated by the Galactic disk. \cite{Nicastro2016b} observed sightlines to Galactic X-ray binaries to measure the contribution of the ISM to the absorption column density of \ovii and showed that to be negligible. Thus it is clear that most of the observed column density of the $z=0$ \ovii absorption lines is from the MW CGM, but the location of the $z=0$ absorbers remains a live question in the literature.  In this paper, we show that a significant amount of hot gas at galactic radii of 116\,kpc and larger is present in an external galaxy (which may be MW-like). This provides strong support for the existence of the extended hot CGM in the Milky Way (and other spiral galaxies). We also note that the average density of the hot gas we measure in this external galaxy between the radii of 116kpc and 200kpc is somewhat smaller than the estimate of average density in the Milky Way CGM of $\rm n_e=2\pm0.6 \times 10^{-4} cm^{-3}$ \citep{Gupta2012}, as expected.  

\subsection{WHIM in the \H18 sightline?}

As noted in \S1, the sightline to the X-ray bright blazar was observed with \chandra LETG-ACIS-S for 500ks to look for the elusive WHIM by \citet{Mathur2003}. They specifically looked for \ovii absorption lines at the redshifts of the six \ovi absorption systems. Subsequently, this sightline was extensively studied to find galaxies close to the intervening absorption systems. \citet{Narayanan2010} reported on the presence of a luminous spiral galaxy at $z=0.225$ at an impact parameter of $116h^{-1}_{71}$kpc from the blazar sightline. They showed that both the z3 and z4 systems arise in the extended halo of this galaxy, probing its CGM. \citet{Kovacs2019} used the same \chandra data of \citet{Mathur2003} to look for the WHIM by stacking the spectra at the redshifts of 17 intervening \lya absorption system. They reported a ``statistically significant'' ($3.3\sigma$) detection of the WHIM.  
We note that the z3 and z4 \ovi systems also have associated \lya absorption \citep{Narayanan2010}, and the $z=0.22489$ \lya absorber associated with the z4 system is one of the 17 \lya absorber systems used as anchors by \citet{Kovacs2019}. This absorber also has the smallest impact parameter of all the \lya absorber systems. Therefore, it is highly likely that the stacking was dominated by the $z=0.22489$ absorber.  

The new 1.2\,Ms data presented in this paper clearly shows \ovii absorption arising in the CGM of the galaxy associated with the z4 system. This, together with the z2 system, were included in the stacking analysis by \citet{Kovacs2019}, dominating the absorption signal. The rest-frame \ovii line EW we measure is $EW=14.3\pm6.9$m\AA, while that reported by \citet{Kovacs2019} is $EW=4.1\pm 1.3$m\AA; these are consistent with each other within $2\sigma$. If the measurement by \citet{Kovacs2019} is correct, the ``true'' rest frame EW will have to be $\sim1.5\sigma$ lower than our best-fit vale. 
In any case, the \ovii absorption arises from within the halo radius of a spiral galaxy at $z=0.225$, therefore it is more accurately characterised as from the CGM of this galaxy. It is unlikely that it is from the WHIM as claimed by \citet{Kovacs2019}.

\begin{figure}
	\includegraphics[width=\columnwidth]{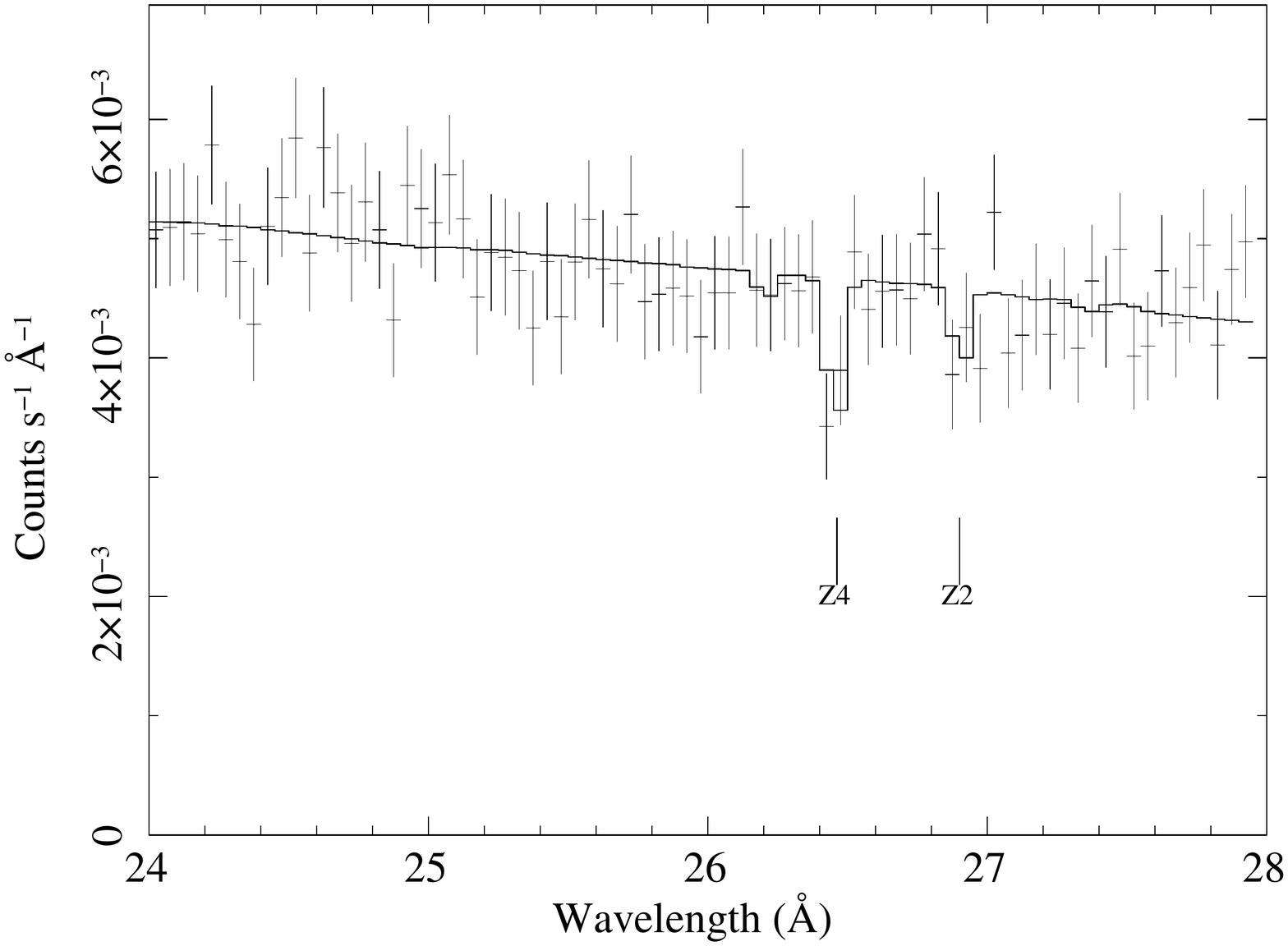}
    \caption{\chandra LETG-HRC-S spectrum in the 24 to 28 \AA\ range. The locations of \ovii K$\alpha$ line at z2 and z4 are marked.}
    \label{fig:allLines}
\end{figure}

\begin{figure}
	\includegraphics[width=\columnwidth]{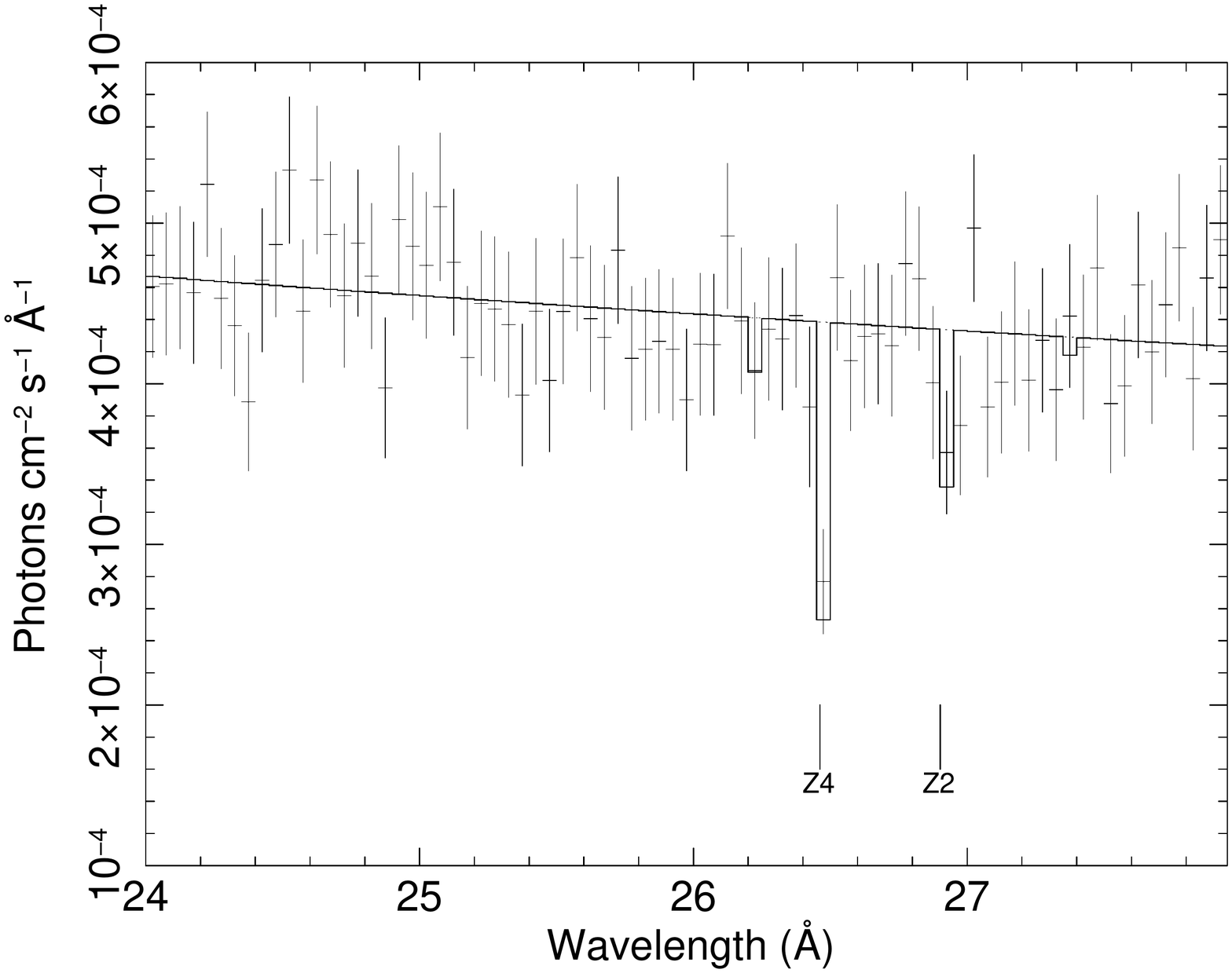}
    \caption{Same as figure 1, but the unfolded spectrum is plotted here.  }
    \label{fig:allLines}
\end{figure}

\begin{figure}
    \includegraphics[width=\columnwidth]{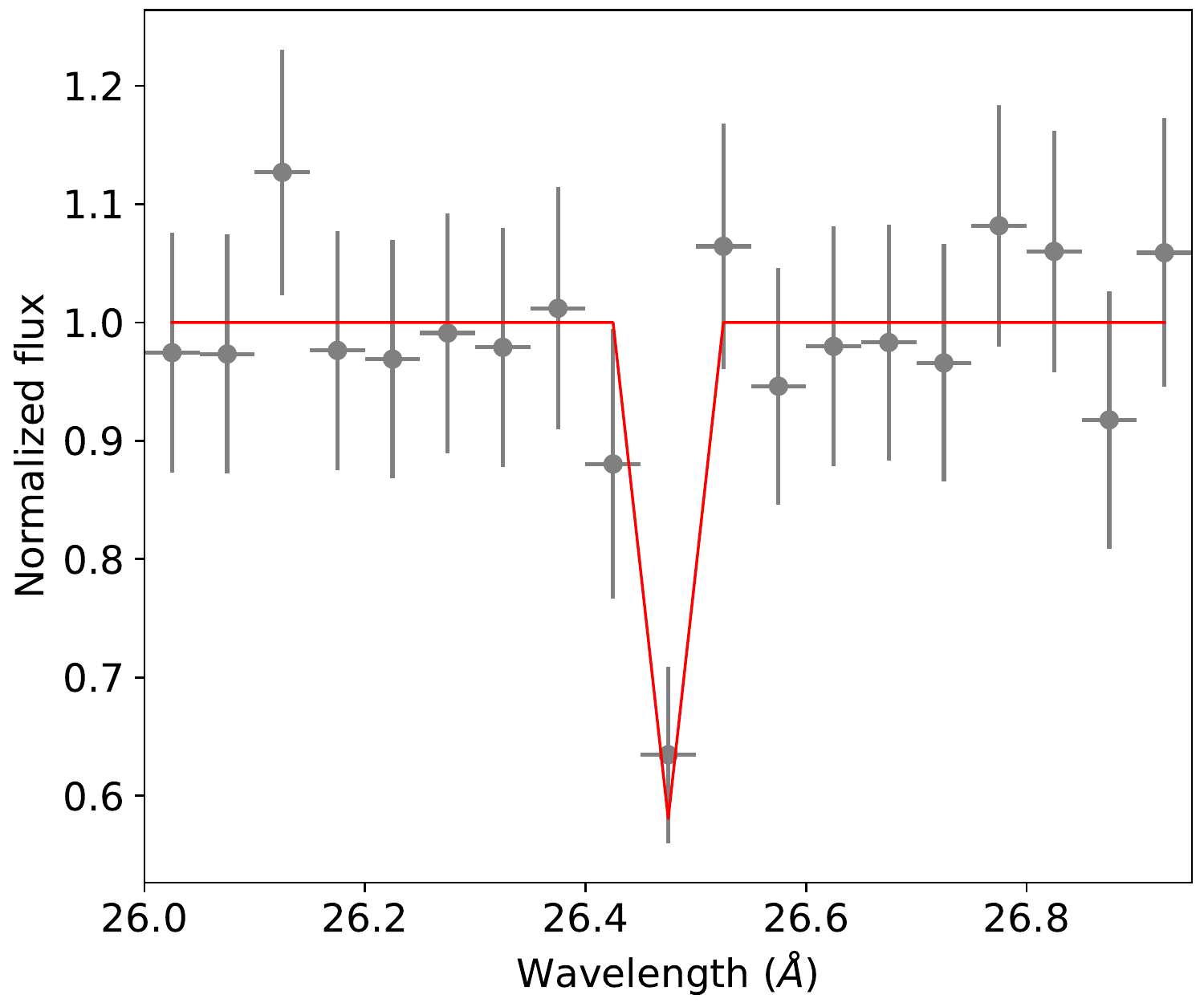}
    \caption{Normalized spectrum between 26\AA\ and 27\AA. Once again, we see the \ovii $K\alpha$ line redshifted to $26.4618$\AA\ at the redshift of the $z=0.225$ galaxy. }
    \label{fig:allLines}
\end{figure}

\section{Conclusions}

We analyzed the 1.2Ms of \chandra LETG-HRC-S data in the sightline to \H18 probing the CGM of an intervening galaxy at $z\approx 0.225$, corresponding to known \ovi absorption systems. We detect the \ovii K$\alpha$ absorption at the same redshift with high confidence, tracing the hot CGM of the galaxy. The presence of such hot CGM was predicted by \citet{Narayanan2010} based on the properties of the UV absorbers. In addition to the $z\approx 0.167$ galaxy in the sightline to PKS$0405-123$ and the ``average'' galaxy detected by stacking analysis of Lyman limit systems (\S 1), this is the 
only other external galaxy in which the hot CGM is detected with absorption line spectroscopy. We argue that the previously reported discovery of the \ovii absorption line from the WHIM in the sightline to \H18 is in fact from the CGM of a luminous spiral galaxy, not from the WHIM. 

\section*{Acknowledgements}

SM is grateful for the grant provided by the National Aeronautics and Space Administration through Chandra Award Number AR0-21016X issued by the Chandra X-ray Center, which is operated by the Smithsonian Astrophysical Observatory for and on behalf of the National Aeronautics Space Administration under contract NAS8-03060. S.M. is also grateful for the NASA ADAP grant 80NSSC22K1121. S.D. acknowledges support from the KIPAC Fellowship of Kavli Institute for Particle Astrophysics and Cosmology, Stanford University. AG gratefully acknowledges support through the NASA ADAP grant 80NSSC18K0419. YK acknowledges support from UNAM PAPIIT grant IN102023. 

\section*{Data Availability}

The scientific results reported in this article are based on
observations made by the Chandra X-ray Observatory (all the ObsIDs of Sequence Numbers: 601449, 601495, 700376). All the data are publicly available in the Chandra X-ray Center (CXC) archive. This research has made use of the software provided by CXC in the application package CIAO. 



\bibliographystyle{mnras}
\bibliography{reference} 





\bsp	
\label{lastpage}
\end{document}